\documentclass[a4paper]{article}

\usepackage{INTERSPEECH2022}
\usepackage{color,soul}
\usepackage{multirow}
\usepackage{graphicx} %插入图片的宏包
\usepackage{float} %设置图片浮动位置的宏包
\usepackage{subfigure} %插入多图时用子图显示的宏包
\title{Towards Error-Resilient Neural Speech Coding}
\name{Huaying Xue$^1$, Xiulian Peng$^1$, Xue Jiang$^2$, Yan Lu$^1$}
\address{
  $^1$Microsoft Research Asia, Beijing, China,
  $^2$Communication University of China, Beijing, China}
\email{huxue@microsoft.com, xipe@microsoft.com, jiangxhoho@cuc.edu.cn, yanlu@microsoft.com}

\begin{document}

\maketitle
\begin{abstract}
Neural audio coding has shown very promising results recently in the literature to largely outperform traditional codecs but limited attention has been paid on its error resilience. Neural codecs trained considering only source coding tend to be extremely sensitive to channel noises, especially in wireless channels with high error rate. In this paper, we investigate how to elevate the error resilience of neural audio codecs for packet losses that often occur during real-time communications. We propose a feature-domain packet loss concealment algorithm (FD-PLC) for real-time neural speech coding. Specifically, we introduce a self-attention-based module on the received latent features to recover lost frames in the feature domain before the decoder. A hybrid segment-level and frame-level frequency-domain discriminator is employed to guide the network to focus on both the generative quality of lost frames and the continuity with neighbouring frames. Experimental results on several error patterns show that the proposed scheme can achieve better robustness compared with the corresponding error-free and error-resilient baselines. We also show that feature-domain concealment is superior to waveform-domain counterpart as post-processing. 
\end{abstract}

\begin{figure*}[htb] %H为当前位置，!htb为忽略美学标准，htbp为浮动图形
\centering %图片居中
\includegraphics[width=1.0\textwidth]{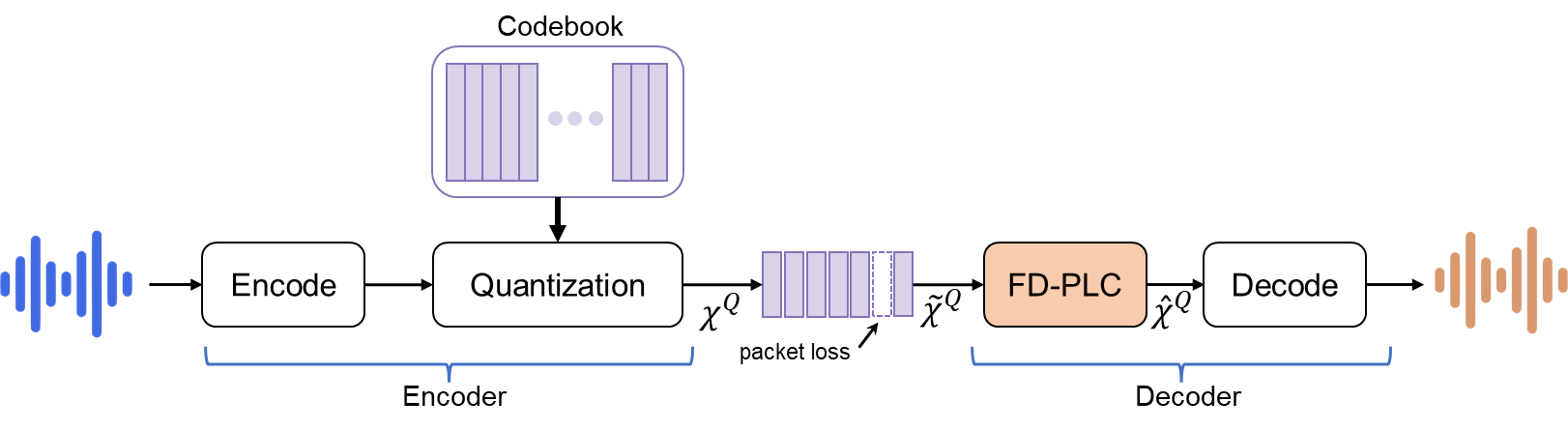} %插入图片，[]中设置图片大小，{}中是图片文件名
\vspace{-0.6cm}
\caption{The overall structure. The FD-PLC module is introduced before decoding to conceal lost quantized features in the feature domain.} %最终文档中希望显示的图片标题
\vspace{-0.6cm}
\label{Fig1} %用于文内引用的标签
\end{figure*}

\section{Introduction}
High-fidelity audio transmission over wireless channels has been increasingly important recently. However, audio packets over the Internet are prone to various types of errors, e.g. random bit errors, packet loss, network congestion and jitters. These errors, if not handled properly, may lead to severe distortion and discontinuity in the received audio. Error resilience is a crucial topic and has been extensively studied in traditional audio coding. Forward error correction (FEC) \cite{hardman1998survey} is a traditional way to protect the compressed bitstream at the sender side. At the same time, modern signal processing based audio coding is typically equipped with a packet loss concealment (PLC) module \cite{stimberg2020waveneteq, stenger1996new} to restore the delayed and missing packets at the receiver side. Numerous studies on adaptive quantizer \cite{simkus2014error}, data partition, unequal error protection \cite{zhou2001error,yang2002error} and PLC algorithms have been proposed to improve the error resilience of the coding. 

In recent years, neural audio/speech coding schemes have shown great vitality in providing extremely high coding efficiency, either by using a strong decoder for recovery from acoustic features \cite{wavcodec,Lyra,sampleRNN,generative} or by end-to-end neural coding \cite{VQ-VAE-wavenet,disentangle,soundstream,cascaded,jiang2022end}. They have demonstrated a high audio quality at a very low bitrate, largely outperformed traditional audio codecs like Opus \cite{valin2012definition}. However, these coding schemes target only at coding efficiency without taking error resilience into account. When there are channel noises, the source coding models are extremely sensitive to channel noises according to our experiments. This paper aims to fill in this gap by investigating how to handle packet losses in a neural coding scheme.

Existing PLC algorithms can be partitioned into two groups, i.e. parametric-domain and waveform-domain PLC. Parametric-domain PLC algorithms aim to predict lost parameters at the codec level, which are used to synthesize the waveform audio. One example is NetEQ \cite{stimberg2020waveneteq}, the standard PLC algorithm in WebRTC that uses Linear Prediction Coefficient (LPC) to estimate the voice and unvoiced components of the signal and interpolates samples as the linear combinations of highly-correlated pitch-periods. Waveform-domain PLC algorithms solve this problem by designing a post-processing step on the decoded waveform. Time-scale modification (TSM) techniques such as Waveform Similarity Overlap-and-Add (WSOLA) \cite{stenger1996new} have been widely used for waveform-domain PLC for its capability of extrapolating audio samples in the time domain with good audio quality. These signal processing based methods yield good quality for short packet losses but tend to produce robot-like artifacts for long burst losses. 

Comparatively, with the significant breakthroughs made in deep learning and generative models, deep-learning based PLC algorithms have been demonstrated to show superior restoration ability recently, especially for long-term packet loss scenario. Most existing deep PLC algorithms are waveform-domain methods introduced to reconstruct lost packets as a post-processing stage. Generally, they can be divided into auto-regressive networks \cite{lin2021time,stimberg2020waveneteq} and generative adversarial networks (GANs) \cite{binkowski2019high, pascual2021adversarial, wang2021temporal}. Auto-regressive methods use recurrent neural networks like LSTM \cite{lin2021time}, WaveRNN \cite{kalchbrenner2018efficient}, WaveNetEQ \cite{stimberg2020waveneteq} as regression models of waveform samples to perform PLC in a real-time setup. These methods usually need a special tuning of the sampling process to generate audio samples after the lost packet and introduce extra delay by feeding output into the network input. In contrast, GAN-based PLC algorithms can generate speech in a frame-in/frame-out manner without any regression \cite{pascual2021adversarial, wang2021temporal}. They typically employ an adversarial training strategy, taking an auto-encoder architecture as the generator and one or more discriminators as a learnable loss function of the restoration task. They have been verified to outperform both the auto-regressive counterparts and the classical methods. One problem of the post-processing based methods is that they are highly dependent on the codec's output. Models trained on uncompressed speech usually degrade a lot when being used directly on the decoded audio without a finetuning or retraining on the codec output. Also, its maximum potential is limited by the codec's quality.

%WaveNetEQ \cite{stimberg2020waveneteq}, a WaveRNN variant, conditioned on the log-mel spectrogram of the past audio, has exceeded its classic counterpart NetEQ, specially for losses beyond 60ms. 

Thanks to the learning capability of neural audio/speech coding, the error resilience could be optimized jointly with source coding to push the boundary further. In our previous work \cite{jiang2022end}, we show an example of this optimization. In this paper, we dig further into this problem and investigate it in a more systematical way. Specifically, we propose a feature-domain PLC (FD-PLC) for neural audio/speech coding, alike the parametric-domain PLC in traditional audio coding. A light-weight attention-based PLC block is introduced to recover lost feature frames at the decoder. This structure is efficient to capture both local and global correlations along the temporal dimension with different attentiveness to lost/non-lost frames. Furthermore, both a spectrogram-based loss at multiple scales and a hybrid segment-level and frame-level adversarial loss are utilized to achieve a natural and temporally coherent reconstruction quality. Taking the end-to-end neural speech coding network \cite{jiang2022end} for real-time communications as a backbone, our experimental results show that the proposed method largely enhances the output quality under several packet loss patterns.

\section{The Proposed Scheme}

A typical neural speech coding network is composed of an encoder, a vector quantizer and a decoder, as illustrated in Fig. \ref{Fig1}. A single channel recording \(x\in\mathbb{R}^L\) is mapped to a sequence of embeddings \(X^S\in\mathbb{R}^{T\times1\times C}\) with low latency. Then the vector quantizer discretizes the embeddings to quantized features \(X^Q\in\mathbb{R}^{T\times1\times C}\) with a set of finite codebooks to meet the target bitrate. Without considering channel losses, the decoder produces a lossy reconstruction \(\hat{x}\in\mathbb{R}^L\) from \(X^Q\). When the channel suffers from packet losses, only the lossy quantized features \(\tilde{X}^Q=\{\tilde{X}^Q_{non\_lost},\tilde{X}^Q_{lost}\}\) are available. The decoder needs to recover \(\hat{x}\) from \(\tilde{X}^Q\) with both quantization noises and packet losses. To facilitate this recovery, we introduce the FD-PLC module just after the inverse quantization in the feature domain. Let \(\hat{X}^Q\) denote the recovered features by FD-PLC. The decoder follows to reconstruct the whole waveform. The whole network is trained on the decoder part to minimize the distortion $D(x, \hat{x})$ at given bitrate and packet loss rate and pattern. Multiple discriminators are designed for adversarial training to produce a natural and temporally-coherent output quality with high fidelity. The following subsections will describe them in detail.   
%The problem can be formalized as a Hidden Markov Model (HMM) \cite{eddy2004hidden}. \(P(H_t\mid O_1,O_2,...O_t)\), in which \(\{O_1,O_2,...,O_t\}\in X^Q_{non\_lost}\) are observable discrete variables at different time \(t\), \(\{H_1,H_2,...,H_t\}\in X^Q_{lost}\) are the hidden discrete variables. The hidden states can only be determined by the current state based on the current and previous observations. Within the context of neural speech coding scheme, we can use data-driven methods to model several important probability distributions in sloving HMM problems, i.e. \(P(H_{t+1}\mid H_t)\) and \(P(O_t\mid H_t)\). Specifically, we design a TSA\_TCM module to predict \(\tilde{X}^Q\) close to \(X^Q\) (see \ref{section:plc}). The model is also trained together with multiple discriminators(see \ref{section:discriminator}), using mix of reconstruction loss to achieve both signal reconstruction fidelity and high perceptual quality.(see \ref{section:training}).
%(todo:add 1 pic)

 \subsection{Backbone network}
 We take the low-latency neural speech coding network TFNet in our previous work \cite{jiang2022end} as the backbone. It takes the time-frequency spectrum with a 20ms window and a 5ms hop length as input with power-law compressed normalization on the magnitude. The encoder and decoder are composed of causal convolutions and deconvolutions for capturing frequency dependencies and two kinds of causal temporal filtering modules in-between them, i.e. dilated temporal convolution module (TCM) and group-wise gated recurrent unit (G-GRU) for capturing temporal dependencies. The two temporal filtering modules are organized in an interleaved way to efficiently extract both local and long-term temporal dependencies. For vector quantization, the latent embeddings \(X^S\in\mathbb{R}^{T\times1\times C}\) are split into \(N\) groups. For each group, it uses an independent codebook containing \(S\) codewords. At the target bitrate of 6kbps, we combine 4 overlapped frames (corresponding to 20ms new data) into a vector for quantization. 
 
 \subsection{FD-PLC module} \label{section:plc}
\begin{figure}[bh]
\centering %图片居中
\includegraphics[width=0.5\textwidth]{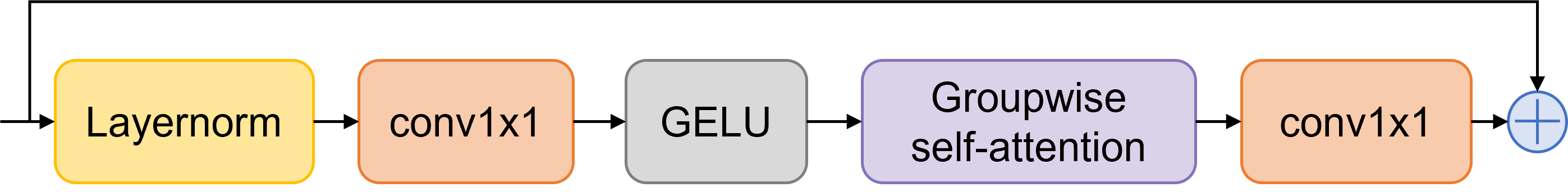} 
\caption{Group-wise temporal self-attention module (G-TSA).}
\vspace{-0.4cm}
\label{Fig2}
\end{figure}

 The FD-PLC module is composed of two kinds of causal modules stacked together, i.e, a group-wise temporal self-attention block (G-TSA) and a TCM module. The G-TSA block is similar to the multi-head self-attention block (MHSA) in transformer \cite{vaswani2017attention} and we turn it into a causal operation along the temporal dimension using a window of \(N\) frames, i.e. each frame only has access to \(N\) past frames without any look-ahead. It captures different attentiveness to different frames thus provides temporal adaptation to the content and the loss/non-loss properties. As shown in Figure \ref{Fig2}, we use a pre-norm residual unit similar to that in \cite{wang2019learning} for the G-TSA block, with two convolutional layers with a kernel size of \(1 \times1\) to reduce and increase the dimensions. The TCM block is similar to that used in the backbone network but with layer normalization preceding the convolutions and GELU as the activation function. Several TCM blocks with increasing dilation rates are stacked together to form a large TCM module with a large receptive field. We use G-TSA to extract the local correlations which is more important to the current frame at a fine granularity and the TCM is used to aggregate the G-TSA output features to catch long-term dependencies at a coarse granularity.

 \subsection{Adversarial training} \label{section:discriminator}
 Adversarial training is widely used for restoration tasks to achieve good reconstruction audio quality. We employ two frequency-domain discriminators with discrimination capability at different granularity. They both take the magnitude spectrum, with a window length of \(W=20ms\) and a hop length of \(H=10ms\) as the input. The first is a segment-level discriminator used to judge the overall quality of an audio clip. It consists of four convolutional layers with a kernel size of (3, 3) and a stride of (2, 2), followed by normalization and Leaky Relu activation. The number of channels is progressively increased to 64 with the depth of network. Finally, a fully connected layer is used to aggregate all channels into one and we do an average pooling on both the (down-sampled) time and frequency dimensions to get just one logit at the output. The second discriminator targets at frame-level discrimination. For this purpose, we use a kernel size of (2, 5) with a stride of (1, 2) in convolution blocks so as to down-sample frequency bins while keeping the temporal resolution. Only frequency-dimension average pooling is used to obtain a 1-dimensional logits as the output. In both discriminators, we use spectral normalization \cite{miyato2018spectral} for the first convolution block and instance normalization \cite{ulyanov2016instance} for others for stable training. Sigmoid activation is used on the output.

We take the Binary Cross Entropy (BCE) loss generally used for GAN \cite{goodfellow2014generative}, i.e. 
\begin{equation}\label{(4)}
\max\limits_{D}L_{adv}\left(D\right)=\mathbb{E}_x\left[log\left(D\left(x\right)\right)\right]+\mathbb{E}_x\left[log\left(1-D\left(G\left(x\right)\right)\right)\right],
 \end{equation}
 \begin{equation}\label{(5)}
\min\limits_{G}L_{adv}\left(G\right)=\mathbb{E}_x\left[log\left(1-D\left(G\left(x\right)\right)\right)\right].
 \end{equation}
 We also use a feature matching loss \cite{kumar2019melgan} as an additional constraint for the generator, which is given by
 \vspace{-0.2cm}
  \begin{equation}\label{(6)}
L_{fm}(G)=\mathbb{E}_x\left[\sum\limits_{i=1}^T\dfrac{1}{N_i}\left\|D^i(x)-D^i(G(x))\right\|_1\right],
\vspace{-0.1cm}
 \end{equation}
 where \(T\) denotes the number of layers in the discriminator that are used for the feature loss. \(D^i\) and \(N^i\) denote the features and feature sizes in the \(i\)-th layer of the discriminator, respectively.
 
 \subsection{Training objectives}\label{section:training}
 The neural coding network works as the generator in adversarial training. We use a combination of several loss terms in guiding it towards a good decoded audio quality as follows
 \begin{equation}\label{(7)}
  \begin{split}
        L_{G}=\lambda_{plc}L_{plc}(G)+\lambda_{bin}L_{bin}(G)+\lambda_{mel}L_{mel}(G)  \\
        +\lambda_{adv}L_{adv}(G)+\lambda_{fm}L_{fm}(G),
  \end{split}
 \end{equation}
where the scalars \(\lambda_{plc},\lambda_{bin},\lambda_{mel},\lambda_{adv},\lambda_{fm}\) are weights to balance different terms and set by \(\lambda_{plc}=1,\lambda_{bin}=1,\lambda_{mel}=0.25,\lambda_{adv}=1e^{-3},\lambda_{fm}=2e^{-5}\) in our implementation. $L_{adv}(G)$ and $L_{fm}(G)$ are adversarial and feature loss terms in Eq. \ref{(5)} and \ref{(6)}.

The first term \(L_{plc}(G)\) adds the supervision on guiding the FD-PLC module to recover the lost frame from \(X^Q\). During training, we add proportional lost frames and non-lost frames to make the data balance. Let $S$ denote this set and the loss term is given by
\vspace{-0.2cm}
\begin{equation}\label{(1)}
\vspace{-0.1cm}
L_{plc}\left(G\right)=\mathbb{E}_x\left[\dfrac{1}{\left|S\right|}\sum\limits_{t\in{S}}\left\|X^Q(t)-\hat{X}^Q(t)\right\|_1\right],
\vspace{-0.1cm}
 \end{equation}
 where $X^Q(t)$ and $\hat{X}^Q(t)$ are $t$-th frame of $X^Q$ and $\hat{X}^Q$, respectively. $\left|S\right|$ denotes the number of frames in $S$. The L1 loss is used to measure the distance between \(\hat{X}^Q\) and \(X^Q\). In our implementation, we take the error-free baseline as the pretrained model for encoder, the codebook and $X^Q$ and train the FD-PLC and the decoder jointly. More ablation studies on the training algorithm could be found in section 4.2. 
 
 The second $L_{bin}(G)$ and the third terms $L_{mel}(G)$ are quality terms at each frequency-bin and mel-band. As shown in \cite{yamamoto2020parallel} that time-frequency distribution can be effectively captured by jointly optimizing multi-resolution spectrum and adversarial loss functions, we use $L_{mel}(G)$ at multiple resolutions to achieve a high perceptual quality, which is given by
 %\vspace{-0.2cm}
 \begin{equation}\label{(3)}
L_{mel}\left(G\right)= \mathbb{E}_{x,\hat{x}}\left[\dfrac{1}{R}\sum\limits_{r=1}^{R}\left\|\phi^r(x)-\phi^r(\hat{x})\right\|_1\right].
%\vspace{-0.1cm}
 \end{equation}
 $\phi^r(\cdot)$ is the $r$-th mel-scale band. To achieve high-fidelity, we also use the frequency-bin wise quality term $L_{bin}(G)$ by
 \vspace{-0.2cm}
 \begin{equation}\label{(2)}
L_{bin}\left(G\right)= \mathbb{E}_{x,\hat{x}}\left[\left\|\psi(x)-\psi(\hat{x})\right\|^2_2\right].
\vspace{-0.1cm}
 \end{equation}
 $\psi(x)$ is the power-law compressed magnitude of $x$. We use L2 distance metric here.
 
\begin{figure*}[htbp]
\centering
\subfigure[]{
\label{Fig3:a}
\includegraphics[width=7cm]{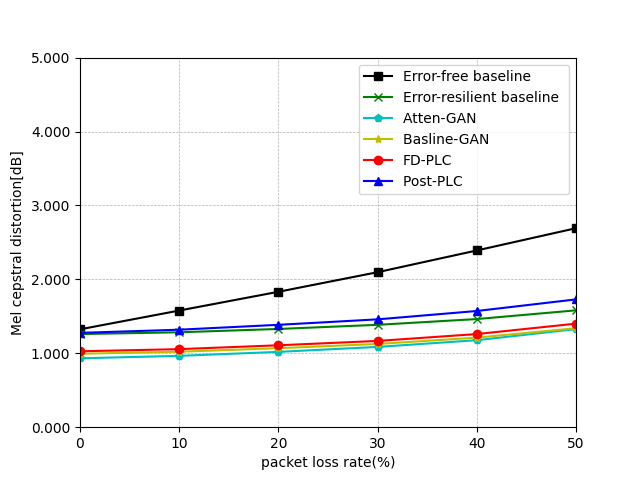}
}
\subfigure[]{
\label{Fig3:b}
\includegraphics[width=7cm]{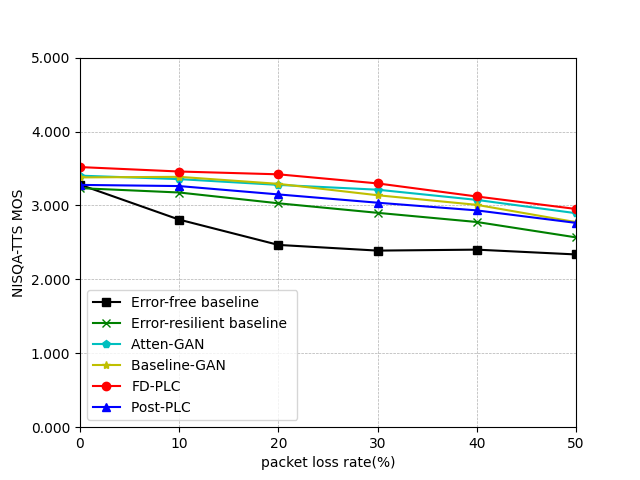}
}
\vspace{-0.4cm}
\caption{Quality comparison MCD(left) and NISQA-TTS(right) at packet loss rate 0\% to 50\% on synthetic traces.}
\label{Fig3}
\end{figure*}

\begin{table*}[h!t]
\centering
\caption{Quality comparison on the blind test set with real traces. Except for MCD, the higher the better.}
%\vspace{-0.2cm}
\begin{tabular}{ |c||c|c|c|c|c|  }
\hline
Scheme&MCD[dB]&PLC-MOS&NISQA-MOS&NISQA-Discontinuity&NISQA-TTS\\
\hline
Error-free baseline                               & 1.859 & 2.461 & 2.799 & 2.844 & 2.85\\
Error-resilient baseline                                  & 1.43 & 3.713 & 3.509 & 3.572 & 2.985\\
Baseline-GAN                        &1.182	&4.159	&3.875	&4.028	&3.121\\
Atten-GAN         &\textbf{1.158}	&4.238	&\textbf{3.978}	&\textbf{4.174}	&3.183\\
Post-PLC          &1.502	&\textbf{4.256}	&3.663	&3.790	&3.076\\
FD-PLC            &1.239	&4.247	&3.957	&4.165	&\textbf{3.325}\\
\hline
\end{tabular}
\label{table:tabel1}
\vspace{-0.2cm}
\end{table*}
 
 \section{Experimental Setup}
 \subsection{Datasets and settings}
 We use the 16khz raw clean speech data from Deep Noise Suppression Challenge at ICASSP 2021\cite{reddy2021icassp}. It includes multilingual speech and emotional clips. For packet loss, we simulate with a random loss rate from \(\{10\%,20\%,30\%,40\%,50\%\}\), whose maximum burst loss length is \(\{60, 80, 120, 160, 220\}\) milliseconds, respectively. Besides, we simulate WLAN packet loss pattern with three-state Markov models\cite{milner2004analysis}. For training, we synthesized 600 hours of data, 100 hours for each loss rate category. For testing, we use the blind test set from Audio Deep Packet Loss Concealment Challenge at INTERSPEECH 2022\cite{diener_2022}. This test set uses real packet loss traces captured in real Microsoft Teams calls. Its maximum burst loss length is up to 1000 milliseconds. We also use another test sets with synthetic traces with a random loss rate from \(10\%\) to \(50\%\) for a deep investigation.
 
 %\subsection{Settings}
 For training, we use the Adam optimizer\cite{kingma2014adam} for both generator and discriminator. A learning rate of \(4e^{-4}\) is employed for generator while for discriminator the learning rate is decayed by \(0.999\) for every epoch with an initial rate of \(4e^{-4}\). The temporal window size $N$ of the FD-PLC module is set to 32 frames.
 %For the experiments, we use one TSA-TCM block for FD-PLC module, with single group, temporal depth of 32 for TSA and 4 TCM blocks with kernel size=5, dilation rate=\{1,2,4,8\} stacked together. So the receptive field is 92 frames. The model size of this module is 6MB. 
 The network is trained for 60 epochs with a batch size of 200.
 
 \subsection{Evaluation metrics}
 We measure signal quality at various packet loss rates while the bitrate is fixed to 6kbps. For a generative task, speech distortion, perceptual quality and naturalness are important factor to measure the quality. To this end, we employ mel cepstral distortion (MCD; in dBs) \cite{kubichek1993mel} to measure speech disortion which focuses on perceptually relevant speech characteristics of the short-term speech spectrum. To measure the overall quality, we use the PLCMOS, the evaluation tool for the PLC Challenge at INTERSPEECH 2022 \cite{diener_2022} and NISQA\cite{mittag2021nisqa}. NISQA is a deep learning framework for speech quality evaluation covering several sources of degradation, including packet losses and audio compression. We use NISQA-MOS to evaluate the overall quality, NISQA-Discontinuity for the audio continuity, and NISQA-TTS for the naturalness of synthesized speech. Our experiments find that they reasonably match our perception when using the same neural audio codec.
 
 \subsection{Baselines for comparison}
 We compare the proposed FD-PLC scheme with several baselines to verify its effectiveness, the error-free baseline, the error-resilient baseline, Baseline-GAN, Atten-GAN, Post-PLC and the FD-PLC as shown in Fig. \ref{Fig3} and Table. \ref{table:tabel1}. The error-free and error-resilience baselines are trained without and with consideration on packet losses. The Baseline-GAN differs from the Error-resilient baseline by adding adversarial training. The Atten-GAN further introduces the FD-PLC module into Baseline-GAN so that it uses the same network as the proposed FD-PLC. It differs from the proposed FD-PLC only in that no $L_{plc}(G)$ loss is used. What's more, the Post-PLC moves the PLC module to waveform domain, acting as a post-processing of Baseline-GAN. The Post-PLC module takes a U-Net structure with causal convolutional encoder and decoder and skip connections between them. TCM and TSA blocks similar to that in FD-PLC are used in-between the encoder and decoder. It is designed with the similar model size as the FD-PLC module but with much larger receptive field.
 
 %To compare with the post-processing scheme, we implement a frequency-domain generative adversarial network. Similar structure as \cite{wang2021temporal} is taken which is based on convolutional encoder-decoder\cite{park2016fully}, with skip connection between encoder layers and decoder layers. For the bottleneck layer, we apply 2 groups of interleaved structure of TSA and TCM. The model size is controlled to be comparable with FD-PLC module, about 6 MB. The receptive field reaches 426 frames in this scheme. Power compressed complex spectrum is also used as the input representation but with 10ms overlap for 20ms frame. The discriminators used are as the same as FD-PLC.
%\begin{figure*}[htb]
%\centering %图片居中
%includegraphics[width=1\textwidth]{LaTeX/visualization_summary_small.png} 
%\caption{Amplitude spectrum generated by different error-resilient schemes. An example of 120ms burst loss. (upper left: Target decompressed at 6Kbps, upper middle: Error-resilient baseline, upper right: Post-PLC, bottom left: Baseline-GAN, bottom middle: Atten-GAN, bottom right: FD-PLC)} 
%\label{Fig4}
%\end{figure*}

\vspace{-0.2cm}
 \section{Results}
  \subsection{Comparison with other schemes}
Fig.\ref{Fig3:a} and \ref{Fig3:b} show the MCD and NISQA-TTS comparison on synthetic traces at different packet loss rates. They measure the signal distortion and the naturalness of the concealed speech, respectively. It can be seen that the Error-free baseline works well when there is no packet losses but the quality drops sharply when there is packet loss, showing its sensitiveness to channel noises. Other error-resilient schemes surpass the error-free baseline not only in loss scenario but also in the non-loss scenario, showing their stronger robustness and restoration capability against the error-free counterpart. Among these error-resilient schemes, except for Error-resilient baseline and Post-PLC, others are comparable in terms of MCD. The proposed FD-PLC scheme achieves relatively lower MCD and highest NISQA-TTS MOS scores consistently over all loss rates. The Post-PLC has much larger MCD, indicating large difference on the signal level. %This is also verified by Fig.\ref{Fig4}, which shows an example of 120ms burst loss where Error-resilient baseline tends to generate blurry results while Post-PLC scheme tends to remove this fake part so as not to bring any audible artifacts. This explains its much higher NISQA-TTS MOS than its preceding error-resilient codec. 
The Atten-GAN also shows promising results, especially at high packet loss rate. Compared with the model without attention module, we observe that more frequency bins are restored since more attention has been put on the lost part. But the content it generates is not as coherent as the proposed scheme. Similar results can be found in Table \ref{table:tabel1} evaluated on the blind test set with real traces. The proposed FD-PLC scheme achieves best NISQA-TTS MOS and comparable results on other metrics with the top one. It also outperforms the Post-PLC on the signal fidelity and perceptual quality. 
%Audio samples could be found at https://fd-plc.github.io/.

 \subsection{Ablation study on training algorithms}
 %\textbf{PLC loss target $X^Q$} Here we investigate different targets $X^Q$ for the PLC loss $L_{plc}(G)$, including features generated by the error-resilient baseline and that by the error-free baseline as utilized in our implementation. We thought this could be a more robust feature which is easier to restore. As shown in Table.\ref{table:tabel2}, the gap between the two is limited, showing that the superiority of FD-PLC against error-resilient and error-free baselines comes from the better restoration capability of the decoder rather than more robust features from the encoder.
 %\begin{table}[h!t]
%\centering
%\vspace{-0.2cm}
%\caption{Quality comparison on different PLC loss targets}
%\vspace{-0.2cm}
%\begin{tabular}{ c|c|c }
%\hline
%Scheme& MCD[dB]&NISQA-TTS\\
%\hline
%Error-resilient baseline                                  & 1.43 &2.985\\
%Features of error-resilient baseline       & 1.3	&	3.125\\
%Proposed         & \textbf{1.239}&	\textbf{3.325}\\
%\hline
%\end{tabular}
%\label{table:tabel2}
%\end{table}

%\footnote{For more results, please visit https://fd-plc.github.io/}

  Here we investigate several training schemes, the end-to-end training, multi-stage training and the proposed. In multi-stage training, the encoder and codebook are pretrained as that in proposed but it trains the FD-PLC module first in decoding and finetunes the decoder after that. As Table. \ref{table:tabel3} shows, end-to-end training is the worst. This is because the model will be confused by the PLC task when the target quantized features are still at a preliminary stage. The proposed joint training of the FD-PLC and the decoder provides more room for the trade-off between packet loss recovery and quantization loss recovery.  
  %\vspace{-0.2cm}
   \begin{table}[h!t]
\centering
\caption{Quality comparison on different training algorithms}
\vspace{-0.2cm}
\begin{tabular}{ c|c|c }
\hline
scheme& MCD[dB]&NISQA-TTS\\
\hline
Error-resilient baseline                                  & 1.43 &2.985\\
End-to-end training       & \textbf{1.222}	&	2.897\\
Multi-stage training         & 1.245	&	3.253\\
Proposed                         & 1.239&	\textbf{3.325}\\
\hline
\end{tabular}
\vspace{-0.2cm}
\label{table:tabel3}
\end{table}

\section{Conclusions}
We propose a feature-domain packet loss concealment algorithm for real-time neural speech coding in this paper. Experimental results show that it could achieve both a better signal fidelity and perceptual quality compared with waveform-domain post-PLC. The proposed self-attention based generative network is able to recover a burst loss with a length of up to 120ms and degrade gracefully with longer burst losses. The proposed FD-PLC module can be easily applied to other neural audio/speech coding networks as well. 
\bibliographystyle{IEEEtran}

\bibliography{mybib}

% \begin{thebibliography}{9}
% \bibitem[1]{Davis80-COP}
%   S.\ B.\ Davis and P.\ Mermelstein,
%   ``Comparison of parametric representation for monosyllabic word recognition in continuously spoken sentences,''
%   \textit{IEEE Transactions on Acoustics, Speech and Signal Processing}, vol.~28, no.~4, pp.~357--366, 1980.
% \bibitem[2]{Rabiner89-ATO}
%   L.\ R.\ Rabiner,
%   ``A tutorial on hidden Markov models and selected applications in speech recognition,''
%   \textit{Proceedings of the IEEE}, vol.~77, no.~2, pp.~257-286, 1989.
% \bibitem[3]{Hastie09-TEO}
%   T.\ Hastie, R.\ Tibshirani, and J.\ Friedman,
%   \textit{The Elements of Statistical Learning -- Data Mining, Inference, and Prediction}.
%   New York: Springer, 2009.
% \bibitem[4]{YourName17-XXX}
%   F.\ Lastname1, F.\ Lastname2, and F.\ Lastname3,
%   ``Title of your INTERSPEECH 2022 publication,''
%   in \textit{Interspeech 2022 -- 23\textsuperscript{rd} Annual Conference of the International Speech Communication Association, September 18-22, Incheon, Korea, Proceedings, Proceedings}, 2022, pp.~100--104.
% \end{thebibliography}

\end{document}